\documentclass[useAMS,usenatbib]{mn2e}
\usepackage[utf8]{inputenc}

\usepackage{graphicx}
\usepackage{amssymb}

\begin{document}

\title[New LBVs in M31]{New Luminous Blue Variables in the Andromeda galaxy}

\author[O.\,Sholukhova et al.] {O.\,Sholukhova,$^{1}$\thanks{E-mail:
olga@sao.ru}
D.\,Bizyaev,$^{2,3}$ S.\,Fabrika, $^{1,4}$ A.\,Sarkisyan,$^{1}$ V.\,Malanushenko,$^{2}$ 
\newauthor A.\,Valeev $^{1,4}$ \\
$^{1}$Special Astrophysical Observatory, Nizhnij Arkhyz, Russia \\
$^{2}$Apache Point Observatory, USA \\
$^{3}$Sternberg Astronomical Institute, Moscow State University, Moscow, Russia \\
$^{4}$Kazan Federal University, Kremlevskaya 18, 420008 Kazan, Russia \\
}
\date{Accepted 2014}

\pagerange{\pageref{firstpage}-\pageref{lastpage}} \pubyear{2014}

\maketitle

\label{firstpage}

\begin{abstract}
We performed spectroscopy of five Luminous Blue Variable (LBV) candidates and two
known LBV stars (AE And and Var\,A-1) in M31. We obtained the same-epoch near-infrared
(NIR) and optical spectra of these stars. The NIR spectra were taken with 
Triplespec spectrograph at the 3.5-m telescope at Apache
Point Observatory, and the optical spectroscopy was done with SCORPIO focal reducer
at the 6-m BTA telescope (SAO RAS). The candidates 
demonstrate typical LBV features in their spectra: broad
and strong hydrogen lines, HeI, FeII, and [FeII] lines. All our candidates show 
photometric variability. We develop a new approach to the LBV parameters estimation based on 
the inherent property of LBVs to change their spectral type at constant bolometric luminosity. 
We compare the spectral energy distributions of the variable stars obtained in two 
or more different states and estimate temperatures, reddening, radii and 
luminosities of the stars using this method. Two considered candidates 
(J004526.62+415006.3 and J004051.59+403303.0) have to be classified as new LBV stars. 
Two more candidates are, apparently, B[e]-supergiants. The nature of one more 
star (J004350.50\,+\,414611.4) is not clear. It does not show obvious 
LBV-like variability and remains an LBV-candidate.   

\end{abstract}


\section{Introduction}

Luminous Blue Variables are massive evolved stars of the highest luminosity. Atmospheres of the stars 
may be highly unstable at the stages when hydrogen is exhausted. Change in ionization state of the most 
abundant elements determines changes in the gas opacity and in the mass loss rate, thus it 
creates a variety of LBV observing manifestations. At the maximum visual brightness LBVs 
show A-F supergiant spectrum, while at the minimum the same star can have a WNL spectrum. 
The relation between LBVs and B[e] supergiants (B[e]SG) is still unclear \cite[]{Kraus2014}: 
they are comparable in luminosities, and similar in spectrum when LBVs are in their hot phase, 
but B[e]SGs do not change their brightness significantly. 

The standard evolution theory does not predict the existence of LBVs. There are 
two possible explanations for LBVs. In the first one
\cite[]{HumphreysDavidson1994, MaederMenet2000} it is considered that 
LBV is a transition stage from the Main Sequence to Wolf-Rayet (WR) stars. The 
second scenario assumes \cite[]{Meynet2011} that LBV is a final stage of high mass 
star's life before Supernova (SN) explosion. An assumption also was made by 
\cite{Smith2014} that LBV can be a product of stellar evolution in binary systems. 

Recent observations of gamma-ray bursts and Supernovae gave an evidence that 
the SN shock travels through an extended stellar wind envelope, as in LBV or WR. 
It was shown than stars can explode as SN at both 
LBV or WR stages \cite[]{Groh2013a,Groh2013b,Grafener2012}.

LBV is a rare class of stars at a particular stage of evolution.
Like other massive stars, they tend to reside close to the galactic midplane, 
where their study is difficult because of high dust extinctions.
An additional problem is the uncertainty in distance estimation. 
It makes studies of extragalactic LBVs with known distances especially valuable. 
\cite{HumphreysDavidson1994} listed five LBVs (four LBV candidates) in the 
Milky Way and 15 in other Local Group galaxies. Four were identified in M31 and four more
in M33. Since that time, 38 LBVs or LBV candidates in the Milky Way have 
been reported \citep{Vink2012}, 24 in M31, and 37 in M33 \citep{Massey2007}.
We have confirmed 4 LBVs in M33 \cite[]{Fabrika2005,Valeev2009,Valeev2010}. 
Using the Spitzer Space Telescope archival data one more LBV and several 
WNL stars in our Galaxy \cite[]{Gvaramadze2009,Gvaramadze2010a,Gvaramadze2010b}
were found. \cite{Humphreys2014} added two more LBV-candidates to M33 and one more to M31.

\cite{Massey2007} classified LBVs by types form their spectral features: cool, hot, and P Cyg LBVs. 
\cite{Humphreys2014} divided the high luminosity stars by six classes based on their 
spectrophotometric features: LBVs, FeII emission-line stars, 
Of/late-WN stars, hot supergiants, intermediate-type supergiants, and warm hypergiants.
\cite{Genderen2001} selected several optical variability types for LBVs: the variables 
with greater than 0.5 mag amplitude, less than 0.5 mag, ex-/dormant LBVs, and candidates. 
 
Majority of LBVs show an infrared spectral excess, which makes their infrared observations favorable. 
\cite{Oksala2013} and \cite{Kraus2014} proposed a new classification scheme for LBVs based on 
their NIR photometry. Massive stars on the (H-K) - (J-H) diagram 
(JHK-diagram) are subdivided by two groups: B[e]SGs have (H-K) $>$ 0.7 mag, 
and LBVs show (H-K) $<$ 0.4 mag. This simple criterion is attractive, although derived 
from a small sample. 

The sample of LBVs with known infrared spectra is rather small, just a few for 
galactic LBVs \cite[]{Morris1996, Voors2000, Groh2007}.
\cite{Kraus2014} present IR spectra for 4 LBV candidates, one of which 
is studied by us in this paper. They also propose to use CO lines for 
identifying B[e]SGs, and have found two such objects. 
It is important to collect more NIR spectral observations for deriving 
reliable classification of the highest luminosity stars.  

In the paper we present quasi-simultaneous optical and NIR spectroscopy for LBVs 
(Var\,A-1 and AE  And) and LBV candidates (J004051.59 +403303.0, J004350.50 +414611.4, 
J004417.10 +411928.0, J004444.52 +412804.0, J004526.62 +415006.3) in the Andromeda galaxy. 
Further in the paper we will refer to the stars using their first RA coordinate. 
We describe their spectra energy distribution (SED) and 
check their LBV nature. The objects were selected from \cite{Massey2007}. 
Their list consists of 4 known LBVs (AE And, AF And, Var 15, Var\,A-1), 
two objects Ofpe/WN9, and 18 more LBV candidates. We adopt the distance to 
M31 of 752$\pm$27 kpc \citep{Riess2012}.

\section{Observations}

Optical spectroscopy and photometry were performed with SCORPIO focal reducer  
\cite[]{AfanMois2005} at the BTA 6-meter telescope in October 2011 and October 2012. 
The observing log is shown in Table 1. 
The data were reduced using standard IDL data reduction package.

The NIR spectroscopy was conducted with the TripleSpec spectrograph at the 3.5-m ARC 
telescope at the Apache Point Observatory  in September 2011 and October 2012. 
The spectra were reduced using Spextool \cite[]{Cushing2004,Vacca2003}.

The BVR photometry was observed simultaneously with the optical spectroscopy. 
We compare our photometry with the U, B, V, R, I data from \cite{Massey2006}
obtained between October 2000 and October 2001. 
Some stars have additional photometry data points published to date: for 
Var~A-1 and AE~And we used U, B, V, R, \& I photometry from September 1976, and J, H, \& K 
from November 1980 from 
\citet{Humphreys1984}. For J004051.6 
\cite{Berkhuijsen1988} published the U, B, V, R, \& I photometry from August - September 1963. 

The JHK photometry is available from the 2MASS survey \cite[]{2MASS}. 
To watch the variability, we used two versions of the catalog: 2MASS-1
\cite[]{Cutri2003} (December 1998), and 2MASS-2 \cite[]{Cutri2012} (November 2000).
There are also JHK magnitudes obtained in November 1980 by \cite{Humphreys1984} for Var~A-1.
The TripleSpec guider at the 3.5-m telescope provides simultaneous K photometry. 
We estimate the K magnitudes for our objects by the comparison with five
non-variable reference stars in the 5-arcmin guider field (using their 2MASS magnitudes). 

Accuracy of our optical photometry is not worse than
0.05 mag. The Massey's estimates have 0.01 mag uncertainty, and 
those by \cite{Humphreys1984} have 0.05 mag uncertainty (0.1 mag in their IR data). 
Our K estimates have 0.1 mag accuracy, while NIR magnitudes from \cite{Berkhuijsen1988} 
are 0.2 mag accurate, and 
the 2MASS-1 and -2 data have 0.1 mag accuracy. 
Results of the optical and NIR photometry are summarized in Table 2.

\begin{table*}
\begin{center}
\caption{Spectroscopy observing log. The columns show the object name, the date of 
NIR Triplespec observations 
(spectral range 0.95 - 2.46 $\mu$, resolution 5\AA), and the optical 
SCORPIO spectroscopy epochs for the ranges 
of 3500-7200 \AA\, (resolution 10~\AA), 
4000 - 5700 \AA\, (resolution 5.5~\AA), 
and 5700-7500 \AA\, (resolution 5.5~\AA). 
The seeing is shown in parentheses. }
\begin{tabular}{ccccc}
\hline
Object & & Date/seeing (arcsec) & & \\
\hline
  &  TripleSpec & & SCORPIO & \\
\hline
J004051.59 & 10.10.12 (1.0) & 18.09.12 (1.3) & 26.10.11 (1.5) & 03.11.11 (1.3) \\
J004350.50 & 17.10.12 (1.1) & 16.09.12 (1.0) & 04.11.11 (1.4) & 04.11.11 (1.2)\\
J004417.10 & 24.09.11 (1.3) & 16.09.12 (1.0) & 26.10.11 (1.5) & 03.11.11 (1.3) \\
J004444.52 & 28.09.11 (1.1) & 19.07.12 (2.6) & 04.11.11 (1.3) & 26.01.12 (1.5) \\
J004526.62 & 24.09.11 (0.9) & 19.07.12 (2.1) & 26.10.11 (1.3) & 04.11.11 (1.3)\\
Var\,A-1 & 24.09.11 (1.0) & 16.09.12 (1.0) & 26.10.11 (1.3) & 04.11.11 (1.3) \\
AE And & 28.09.11 (1.2) & 16.09.12 (1.0) & 27.01.12 (2.0) & 26.01.12 (1.5) \\
\hline
\end{tabular}
\end{center}
\end{table*}
\section{Results}

\subsection{Spectroscopy and Spectral Energy Distributions}

Fig. 1 shows spectra of seven considered stars in six spectral ranges. The
principal lines are identified and marked in Fig. 1: Balmer, Brackett, and
Paschen lines, as well as HeI, HeII,
FeII, [FeII], and SiII lines. Some of these lines demonstrate P\,Cyg-type
profiles. The spectra are typical for LBV stars.
Object J004417.10 has $^{12}$CO lines in its spectrum, which is typical
for B[e]SGs (an indicator of warm stellar media).

\begin{figure*}
%\epsscale{1.05}
\includegraphics[width=170mm]{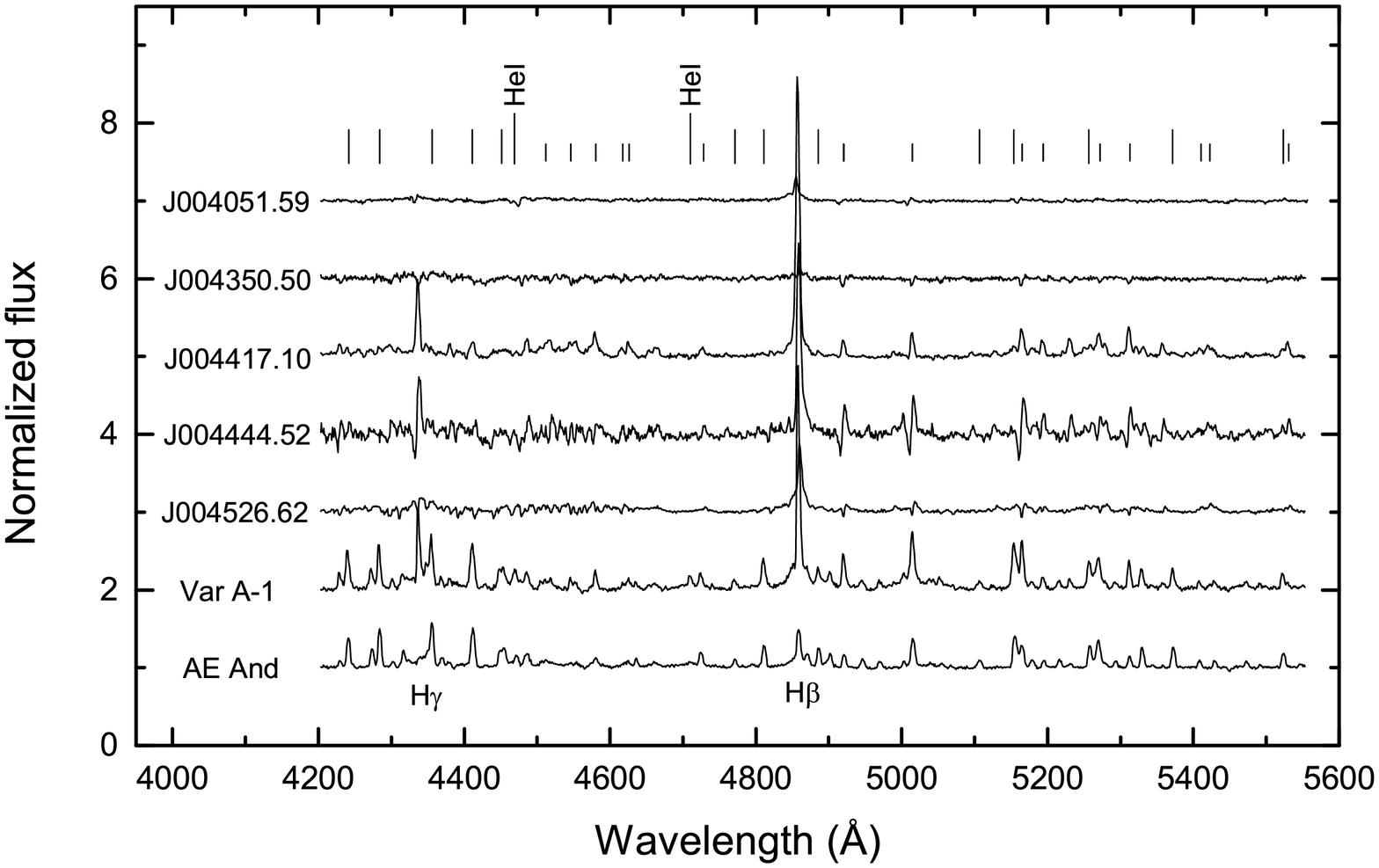}
\includegraphics[width=170mm]{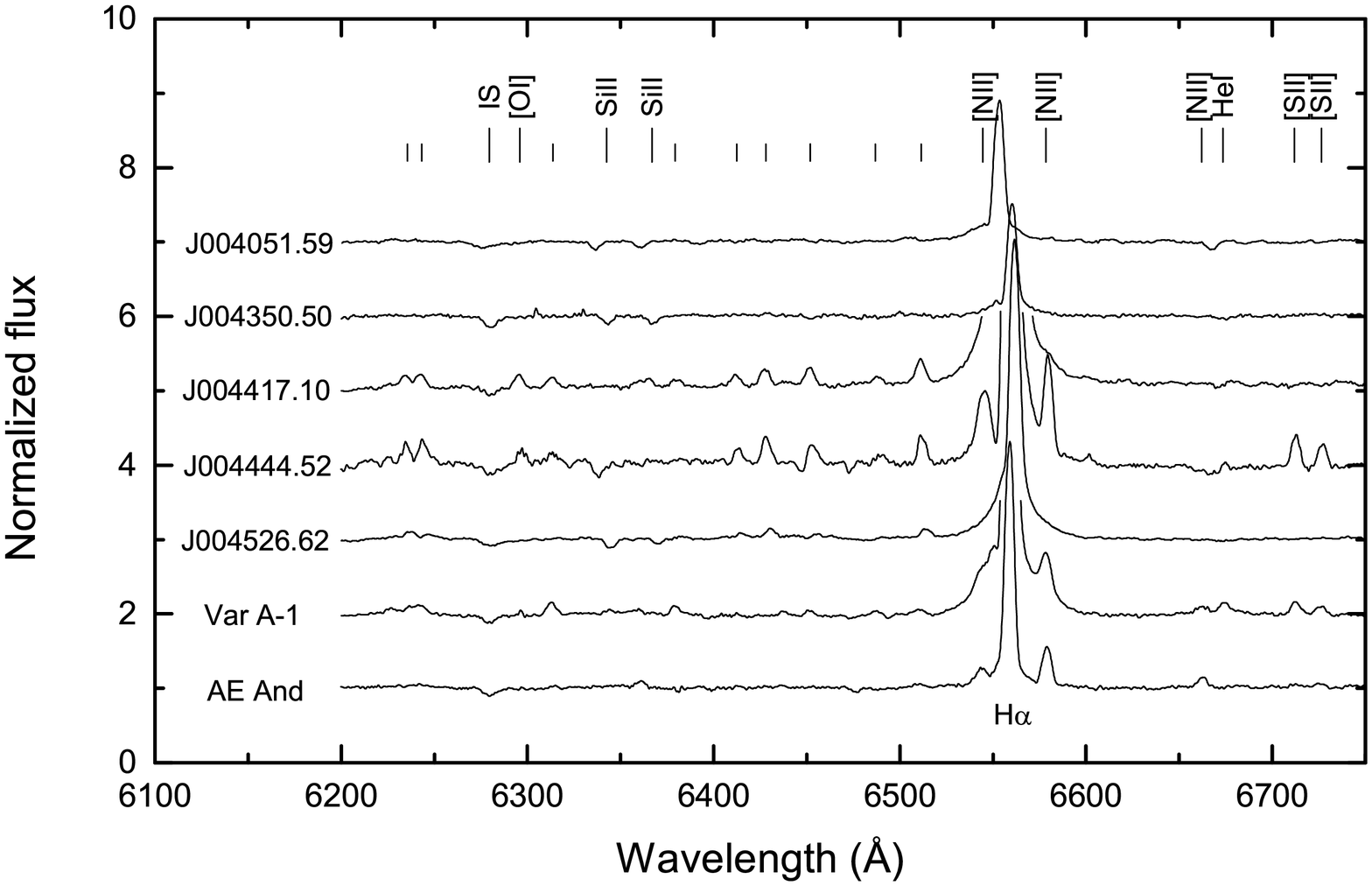}
\caption{The spectra of J004051.59, J004350.50, J004417.10, J004444.52, J004526.62, Var\,A-1, AE And 
in the optical spectral ranges. The principal strongest lines are identified. 
Unsigned short and long tick marks designate the FeII and [FeII] lines, respectively. 
The spectra are left on their original wavelength scale.}
\end{figure*}

\begin{figure*}
\includegraphics[width=170mm]{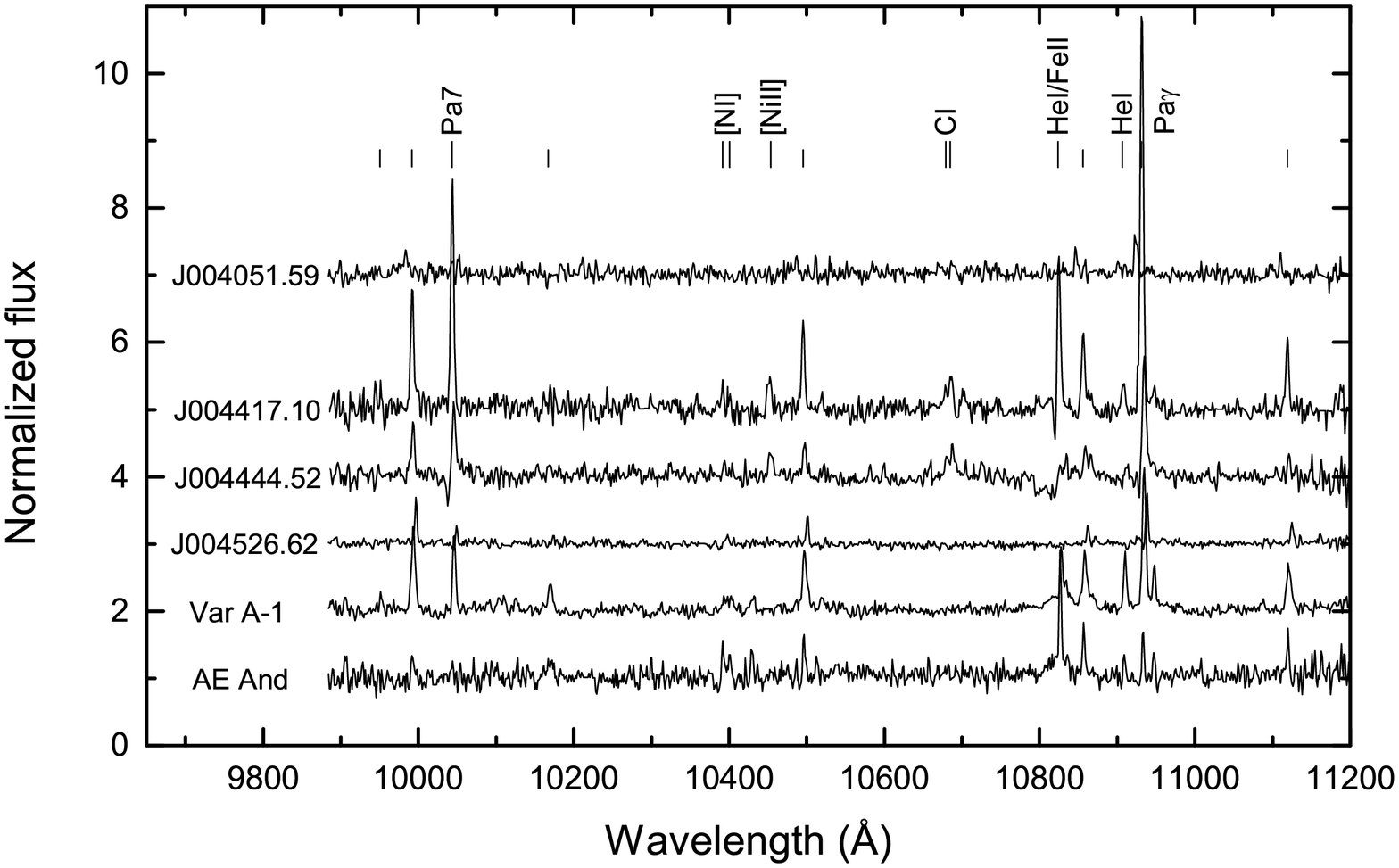}
\includegraphics[width=170mm]{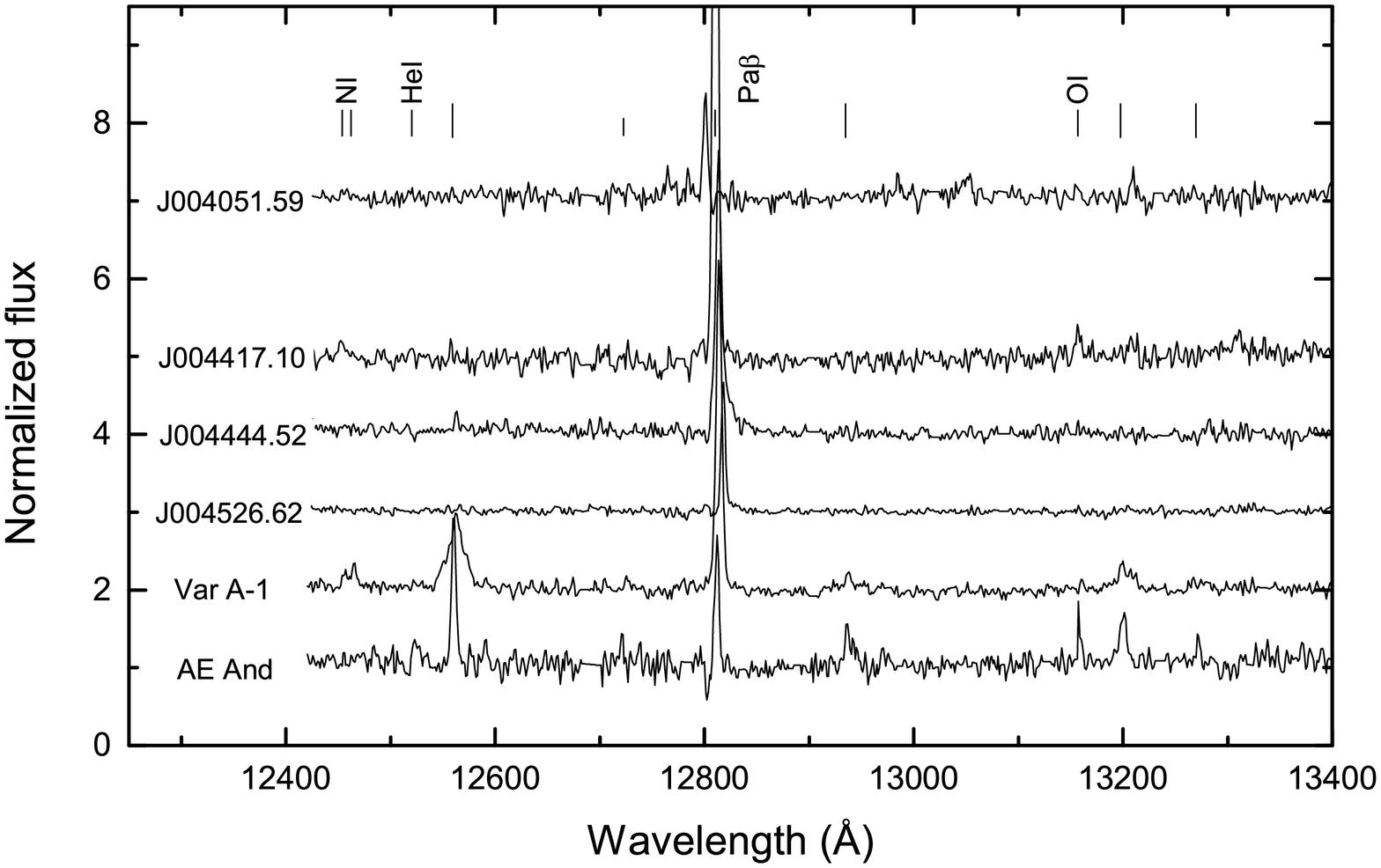}
\caption{The same as in Fig.\,1 but in the J NIR range. The IR spectrum of J004350.50 is not shown. }
\end{figure*}

\begin{figure*}
\includegraphics[width=170mm]{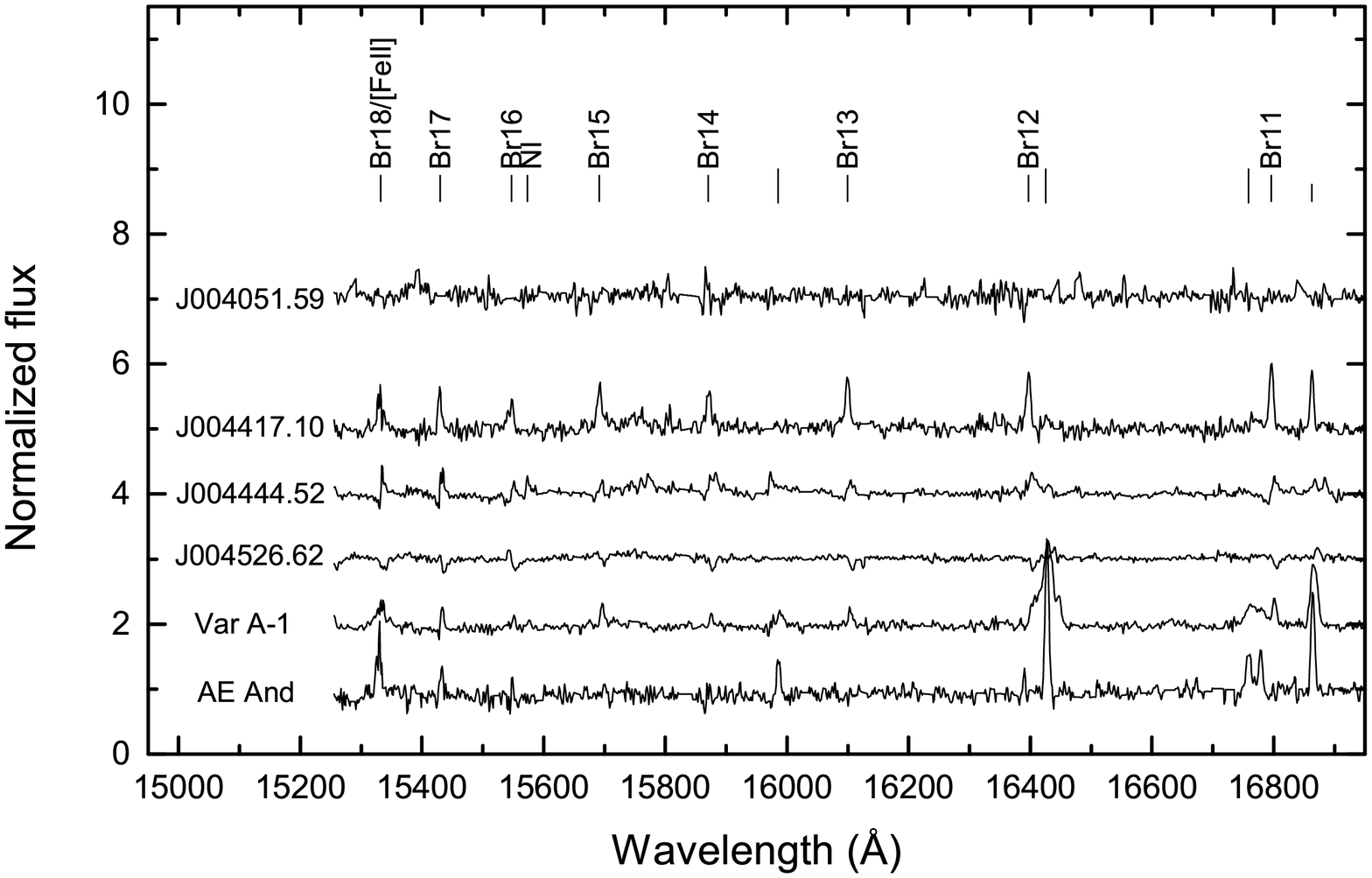}
\includegraphics[width=170mm]{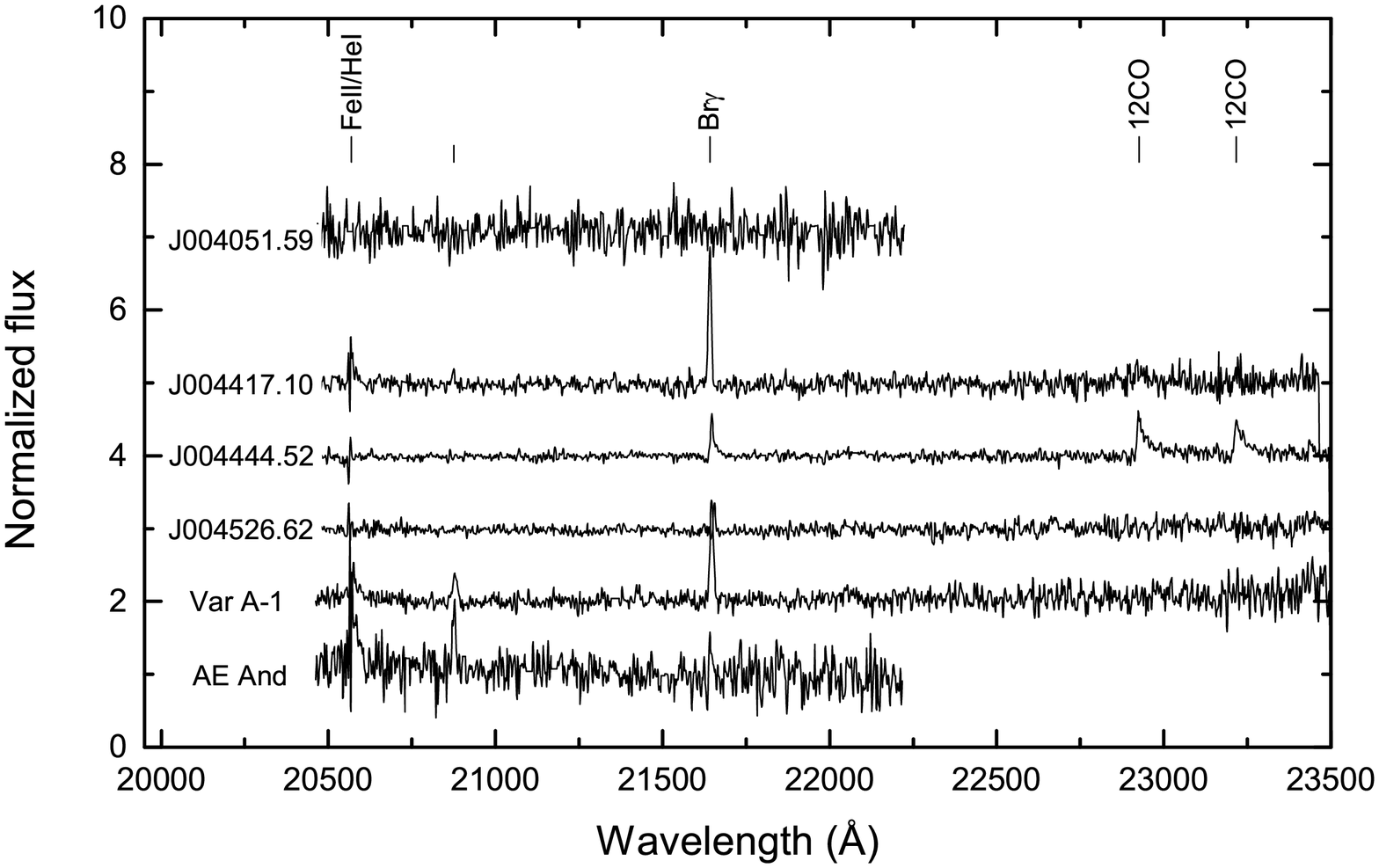}
\caption{The same as in Fig.\,1 but in the H \& K NIR ranges. The IR spectrum of J004350.50 is not shown. }
\end{figure*}

We use the collected photometry (Table 2) and our spectra to study SEDs for our objects. 
To increase the flux calibration accuracy, we tied our calibration to simultaneous 
observations in the V and K bands. Fig. 4 demonstrates our optical spectra 
together with the simultaneous and historical photometric data. 

\begin{table*}
\begin{center}
\caption{The photometric data used in the paper. The columns show the object 
names along with corresponding U, B, V, R, I, J, H, and K magnitudes. 
For each object, the first line shows our B, V, R, and K photometry made 
simultaneously with our spectra. The second line shows the data from 
\protect\cite{Massey2006} obtained between October 2000 and October 2001,
and the J, H, and K from 2MASS-1 and 2MASS-2 \protect\cite[]{Cutri2003, Cutri2012},
obtained in December 1998 and November 2000, respectively.
The third line shows the data from \protect\cite{Berkhuijsen1988} for J004051.6 
(August - September 1963) and from \protect\cite{Humphreys1984} for Var\,A-1 and AE And 
(optical photometry from September 1976, and J, H, \& K from November 1980). 
The data uncertainties are described in the text.}
\begin{tabular}{lcccccccc}
\hline
Object & U & B & V & R & I & J & H & K \\
\hline
J004051.59 & &17.29 & 16.99& 16.76 & & & & 15.96$\pm$0.05\\
            & 16.444 & 17.205& 16.989 & 16.769 & 16.576 & 16.38/16.18& 15.59/15.98& 15.75/15.06 \\
 & 16.93&	17.67&	17.43&17.08&	16.84& & & \\
\hline
J004350.50 & &18.32 &17.73 & 17.19& & & & 15.56$\pm$0.28\\
  &17.986	&18.342	& 17.700	& 17.229	&	16.74	& 16.25/16.21& 15.85/15.80& 15.95/15.71 \\
\hline
J004417.10& &17.37 &17.27 &17.05  & & & & 15.2$\pm$0.28\\
 &16.494	& 17.26	&17.113	&16.78	&16.610&  15.97/16.06 & 15.58/15.55 &14.73/14.63 \\	
\hline
J004444.52 & &19.0 &18.16 & 17.40& & & & 14.4$\pm$0.1\\
& 18.978	&19.062	&18.073	& 17.326	& 16.561& 15.81/15.75 &15.23/15.10 &14.38/14.26 \\
\hline
J004526.62 & &16.82 &16.37 &16.08 & & & & 15.03$\pm$0.2\\
&16.92& 17.662	&17.36	&17.02	&16.920	&15.93/16.62 &15.76/15.84 & 15.13/15.37 \\
\hline
 & & 17.14 & 16.77	& 16.47& & & & 15.5$\pm$0.2\\
Var\,A-1 &16.75	& 17.364 & 17.08 & 16.76 &16.641 & 15.75/16.14& 15.54/16.07 &15.46/15.66 \\
 & 16.13&	16.67	&16.260	&15.76	&15.55& 15.47& 15.24 &15.13 \\
\hline
 & &16.77 &16.57 & 16.31& & & & 15.9$\pm$0.09\\
AE And &16.616	&17.385	&17.373	& 17.242& 17.241 &  16.39/16.90& 15.89/16.64& 15.89/16.33 \\
 & 16.29		&17.1	&17.00	&16.66 &	16.48	& & & \\
\hline
\end{tabular}
\end{center}
\end{table*}

\begin{figure*}
%\epsscale{.80}
\includegraphics[angle=90,scale=0.32]{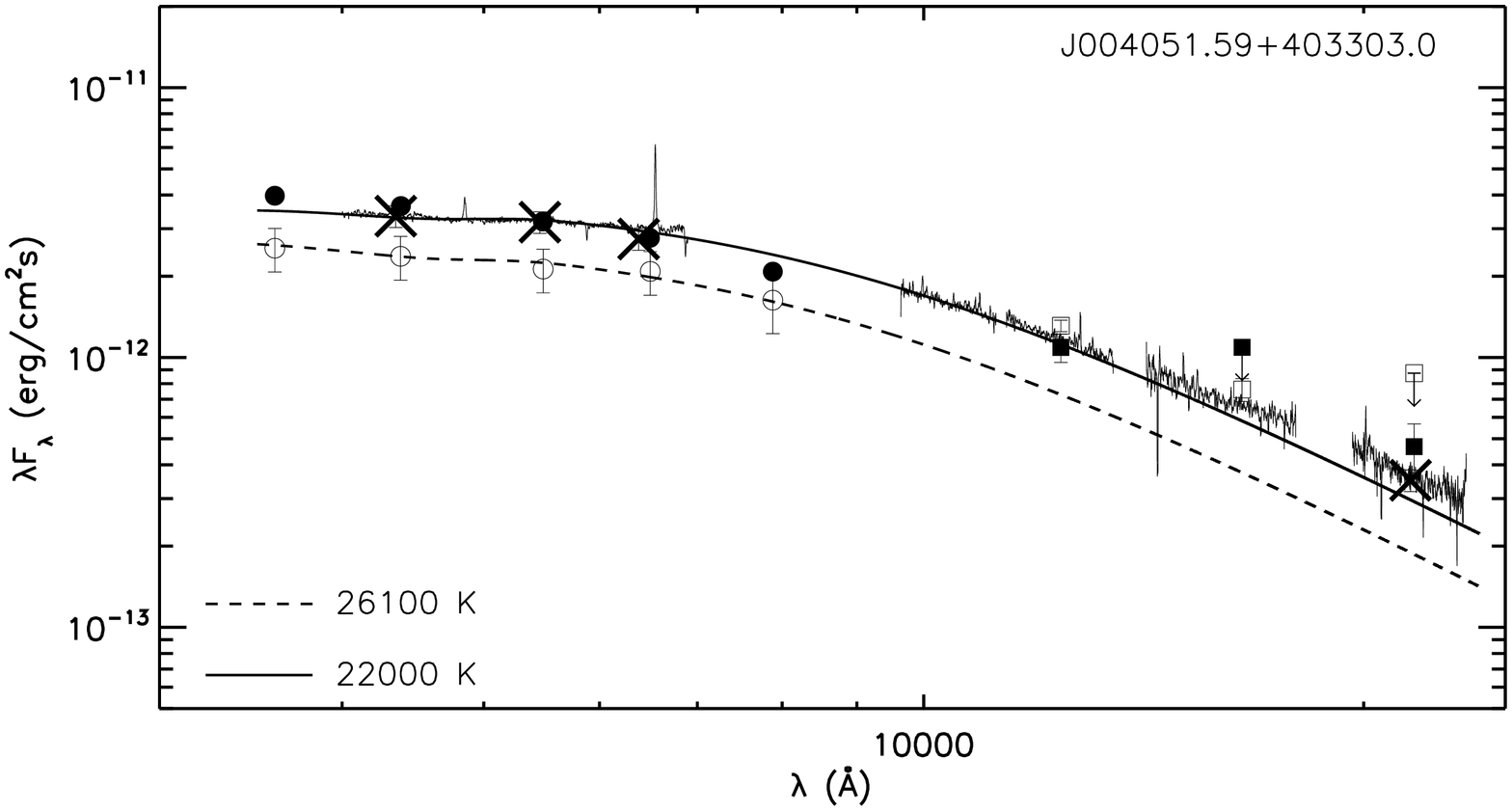}
\includegraphics[angle=90,scale=0.32]{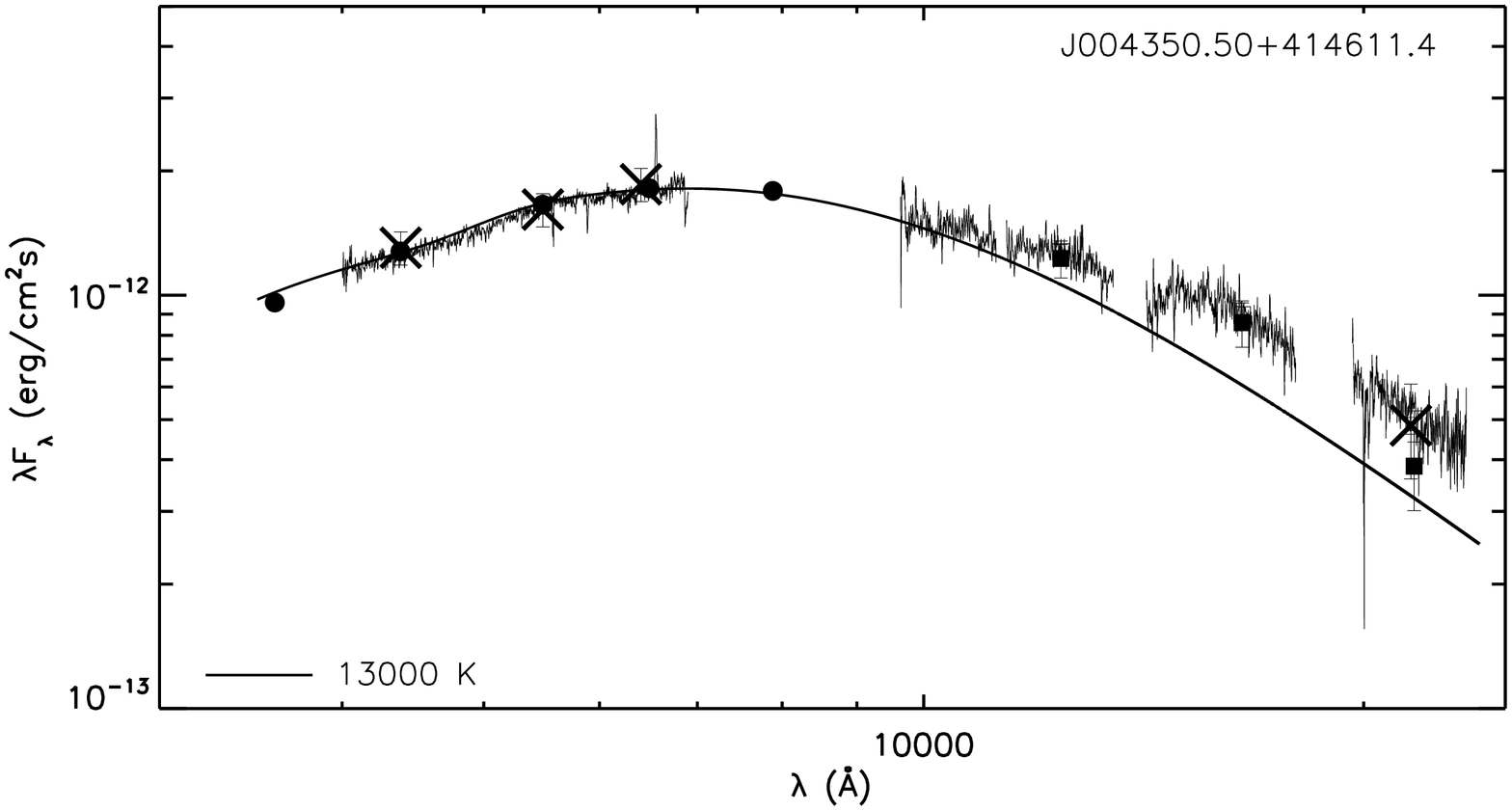}
\includegraphics[angle=90,scale=0.32]{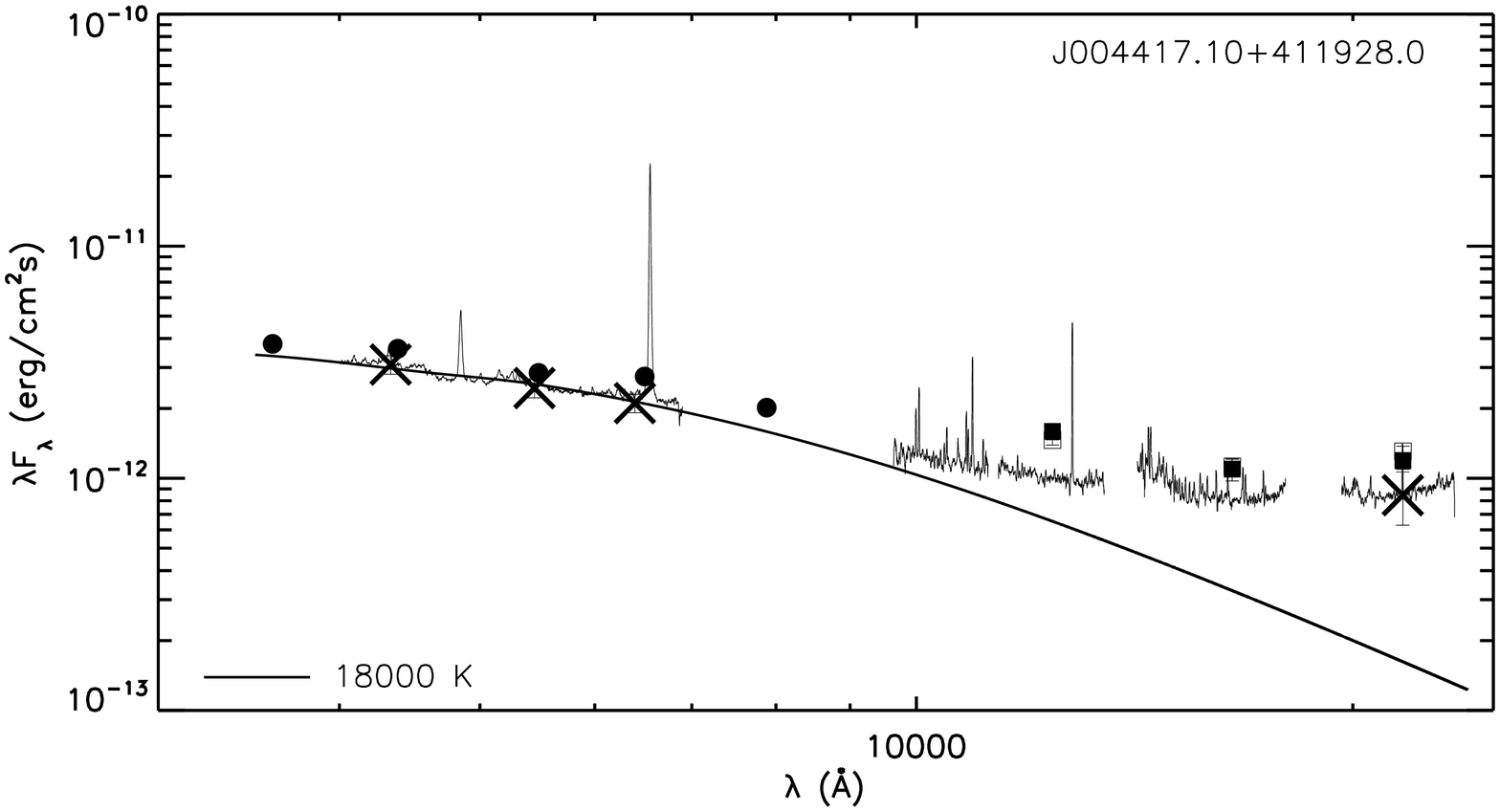}
\includegraphics[angle=90,scale=0.32]{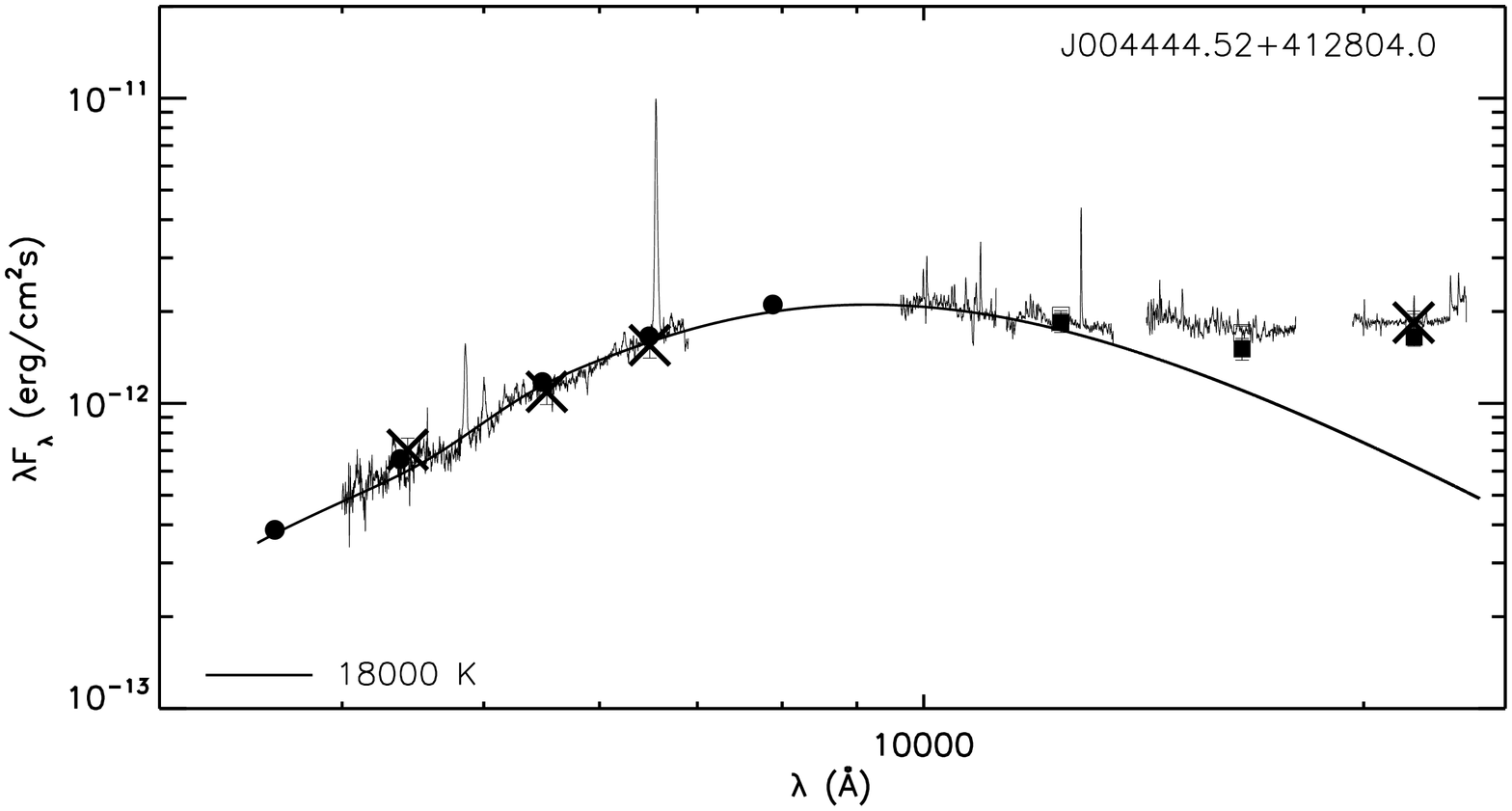}
\includegraphics[angle=90,scale=0.32]{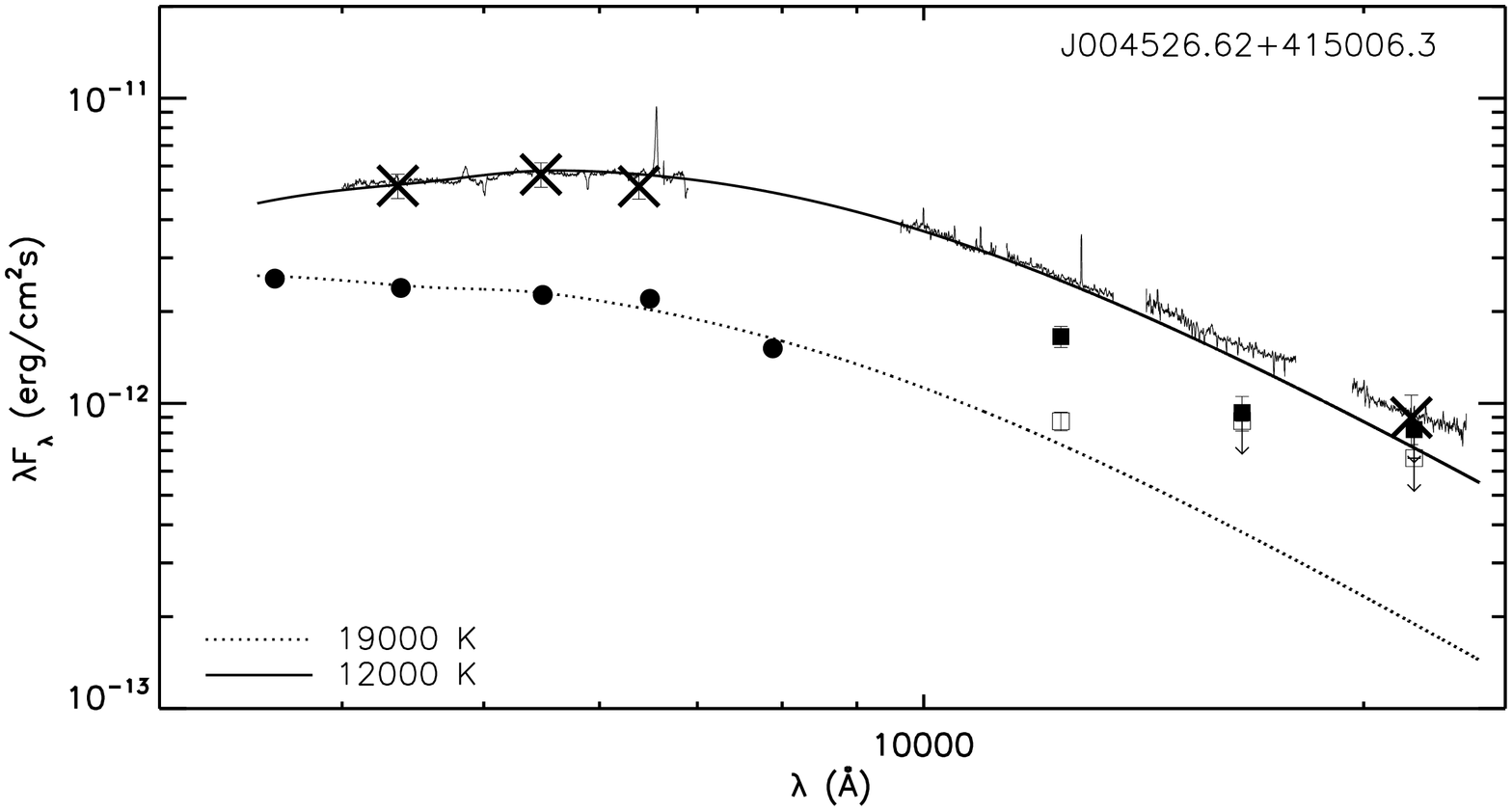}
\includegraphics[angle=90,scale=0.32]{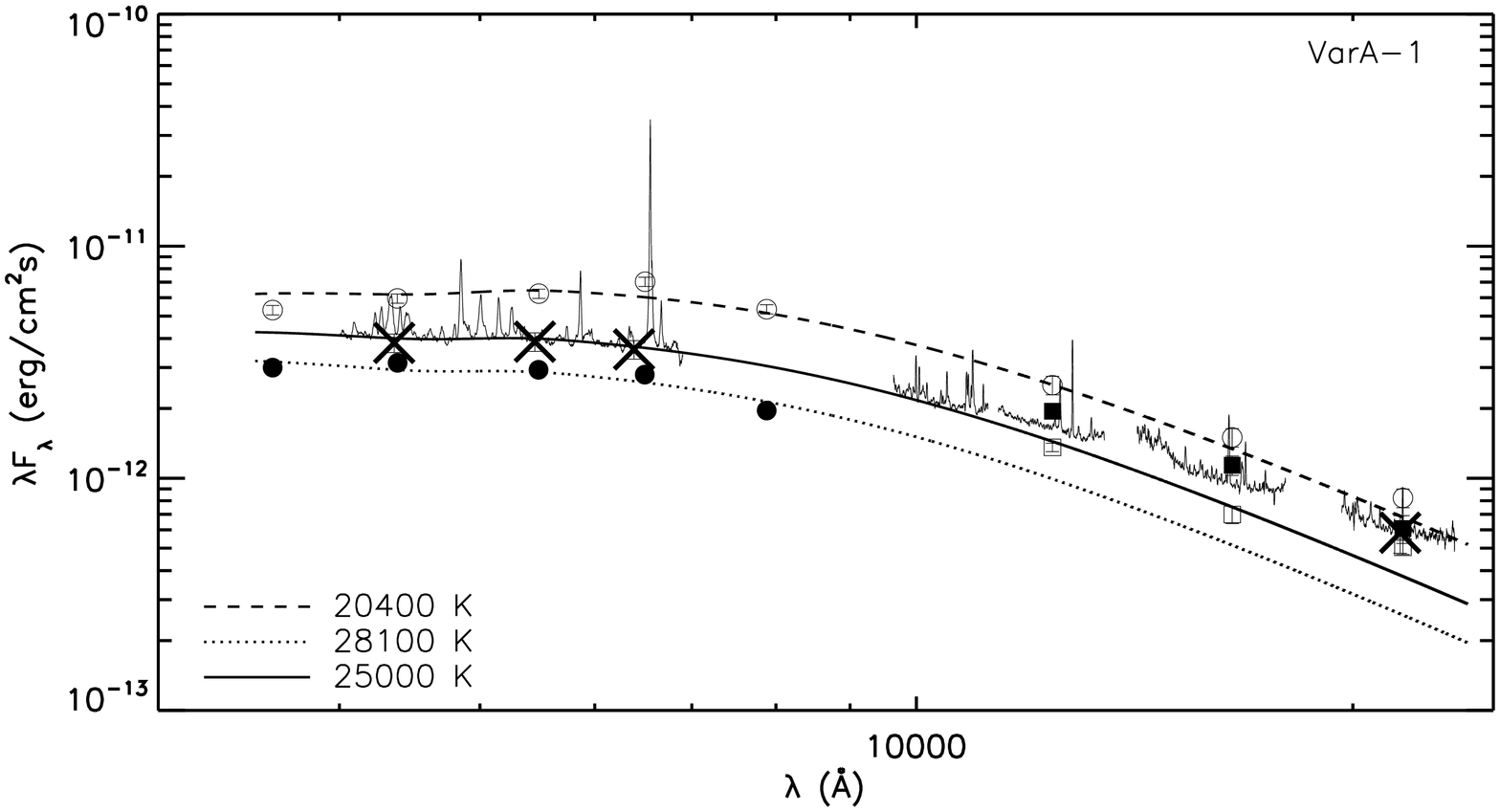}
\includegraphics[angle=90,scale=0.32]{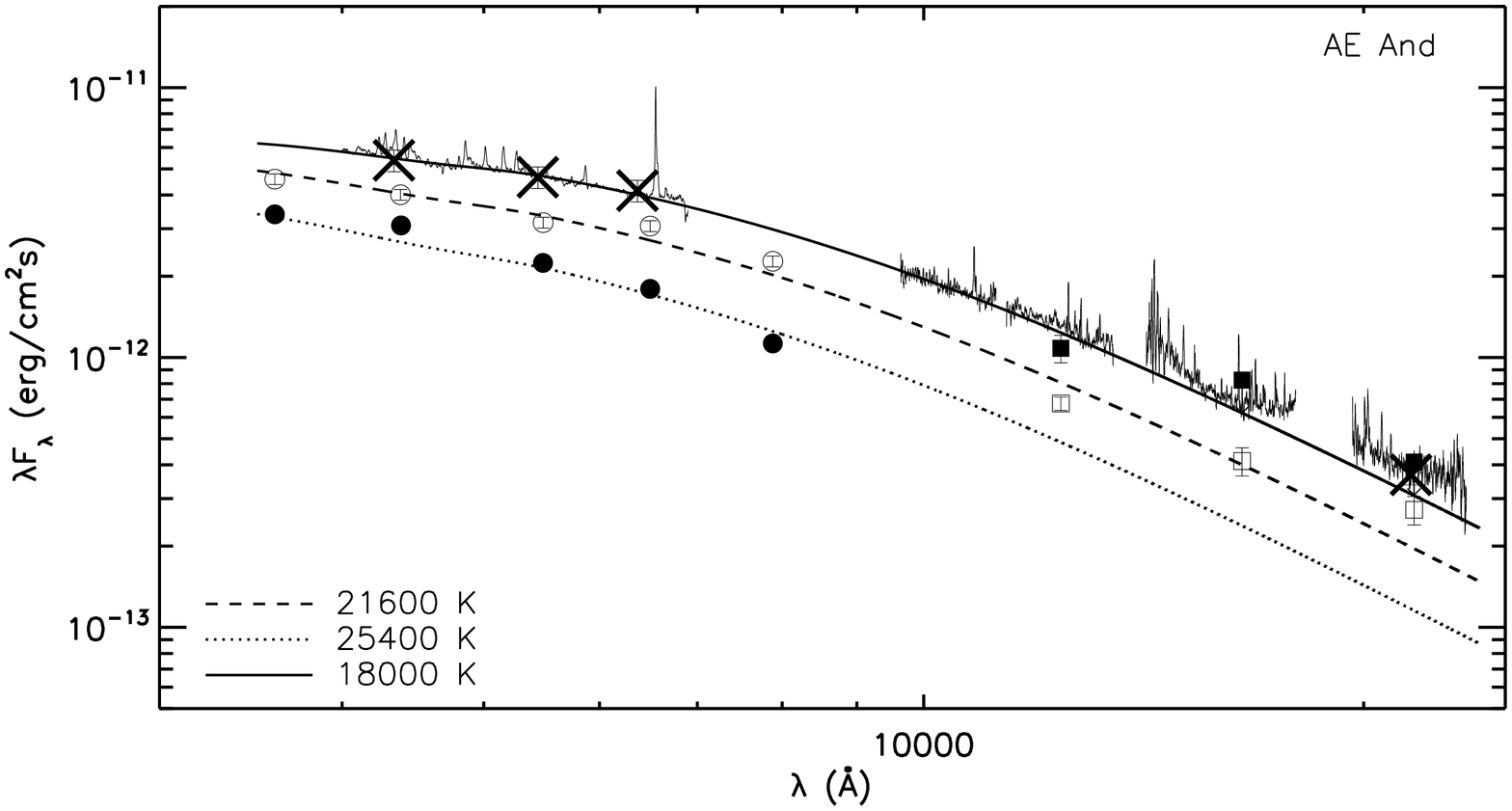}
\caption {The SED modeling. Crosses indicate the B, V, R, K photometry observed simultaneously 
with our spectra, filled circles are the data by \protect\cite{Massey2006},
open circles from \protect\cite{Humphreys1984}, and \protect\cite{Berkhuijsen1988} (Table 2).
Filled and open squares are the data from 2MASS-1 and 2MASS-2 \protect\cite[]{Cutri2003, Cutri2012} 
respectively. 
The curves designate the black body approximation with reddening applied according to Table 1.
The solid curves show our fits to the optical part of our spectra, the
dashed curves --- to the data of \protect\cite{Humphreys1984} and \protect\cite{Berkhuijsen1988}, and the 
short dashes --- to \protect\cite{Massey2006} data.
%The dot dashed curves mark the dust SED fitting. { \bf In all cases we adopted the dust 
%temperature of 1500\,K, except J004444.52 where the temperature is 1300\,K.} 
%The solid curves demonstrate the best-fit sum of the stellar and dust SEDs.
The IR excesses are clearly seen in the stars.
The best-fit temperatures are indicated at the legends in each panel.
}
\end{figure*}

Our spectra allow us to estimate approximate ranges of photospheric temperature
($T_{sp}$ in Table 3).
We used simple criteria of the line relative intensities (HeI, HeII 4686~\AA, 
and FeII) in the spectra \citep{Jacoby1984}.
Next, the optical spectra were fitted with a Planck function taking account 
the extinction with $R_V$ = 3.07 \citep{Fitzpatrick1999}. 

There is a well-known degeneracy between the reddening and temperature,
which makes estimation of the parameters ambiguous. Nevertheless, when the temperature is comparatively low, given the $T_{sp}$ ranges,
we can constrain the extinction $A_V$ rather tightly. We also used the Balmer line ratio H$\alpha$ / H$\beta$ in the case of Var\,A-1 and J004350.50 to estimate the reddening $A_V$ of surrounding nebulas, \citep[see e.g. ][]{Valeev2009}. All estimates mentioned above were used as the 
initial parameters for the further SED fitting.

LBV-type variability allows us to break the degeneracy problem 
when the star gets cooler and brighter, or hotter and dimmer in the optical 
bands \cite[]{HumphreysDavidson1994} with about constant bolometric luminosity.
The variability should be large enough ($\Delta V \gtrsim 0.2$ mag) and rather 
slow (at least a few months, \cite{Sholukhova2011}). 
In this case the constancy of the bolometric luminosity ($\sigma T^4 4\pi R^2 $, 
where R is the stellar radius) is a correct assumption.

We conclude that four object out of seven in our sample show such a slow and large variability. In this case we can fit SED for different data sets 
assuming that the interstellar reddening $A_{V} = const$. Assuming $T^4 R^2= const$, we can estimate the photospheric temperature from known V-magnitude in a new 
state of the star. Fitting the SEDs, we use a black-body approximation for continuum spectra masking strong emission lines and the Balmer jump.  

For J004051.59 and J004526.62 we have data from two epochs, which record two different stellar states, 
and which allow for two temperature solutions. In the case of Var\,A-1 and AE And, we have data 
for three solutions, which enables us to constrain the temperatures T$_{SED}$, $A_V$, and $R$ 
even more precisely. We fit each data set, both photometric and spectroscopic.
The results of the fitting are summarized in Table 3, where we show the reddening $A_V$, 
stellar temperatures, radii, and absolute magnitudes in the V-band and bolometric. 

LBVs and B[e]SGs often show infrared excess that is caused by free-free emission or/and 
thermal dust emission in the near-infrared (JHK) range. The excesses are clearly seen in Fig.\,4.
In two stars (which we classify below as B[e]-supergiants) the IR excesses are very strong.
In our forthcoming paper we will present spectroscopy of more LBVs and LBV candidates in M31. 
In that paper we will analyze IR excesses of these and new stars to have more representative data. 

\subsection{Notes on individual objects}

{\it J004051.59+403303.0} \,
The brightness and color were estimated by 
\cite{Berkhuijsen1988}: in 1963 it had V = 17.43 $\pm$ 0.18 ; 
\cite{Magnier1992}: in 1990 it had V = 17.33, (B - V) = 0.09.
Our estimate, V = 16.99 $\pm$ 0.05, is well consistent with that by \cite{Massey2006}. 
The collected photometry suggests a gradual increasing of stellar optical brightness. 
Our optical spectrum is similar to that published by \cite{Massey2006}: 
it has absorption HeI, FeII and SiII 6347, 6371~\AA\, lines. The brightest FeII lines have P\,Cyg profiles.
There are weak [CaII] 7291, 7323~\AA\, emission lines. 
The data by \cite{Berkhuijsen1988} agree well with a hotter state of this object
(Fig. 4 and Table 3), and they are confirmed by the \cite{Magnier1992} data. 
The photometric variability of this object between its two states ($\Delta$ U=0.49, 
$\Delta$ B=0.46) is LBV-like. Given the spectral and photometric variability, 
and the location on the JHK diagram \cite[]{Kraus2014}, we conclude that J004051.59 
may belong to the LBV class.

\noindent
{\it J004350.50+414611.4}\, 
In the DIRECT project \cite[1996-1997,][]{Stanek1999} this star is 
classified as a Miscellaneous Variable \cite[]{Bonanos2003}. Its 
variability was also detected by \cite{Vilardell2006} ($\Delta$B = $\Delta$V = 0.16 
with uncertainty under 0.01 mag). Our spectra are similar to 
those by \cite{Massey2007}. They have broad Balmer emission lines. FeII lines have 
P Cyg profiles, but HeI and SiII lines are in the absorption. 
The interstellar reddening estimated using the emission from surrounding HII 
regions $A_V$ = 1.6 agrees well with estimates made from our SED fitting. 
\cite{Humphreys2014} classified this star as an intermediate-type supergiant (A5I). 
The location on the JHK diagram \cite[]{Kraus2014} suggests that 
J004350.50 may belong to the LBV class. The star is similar to LBV by 
its spectra and luminosity. However, its relatively small brightness
variations resemble those of $\alpha$\,Cyg type, which may be caused
by fluctuations in the stellar wind. The brightness variation amplitude seen up to date 
does not allow us to classify the star as an LBV. The star may be a dormant LBV, 
however its nature is not clear yet. It remains an LBV-candidate. 
We need to look more additional criteria (e.g. to study helium content in its wind) 
to verify the star's LBV classification.

\noindent
{\it J004417.10+411928.0} \,
It was referred to as a variable with the amplitude of 0.15 in the V band by 
\cite{Mochejska2001}. We notice its spectral variability. Our spectra show 
the line HeI 5876 \AA\, while it was not seen in September 1995 by \cite{King1998}. 
In addition, FeII emission lines got significantly brighter 
that in the spectra from \cite{Massey2007}. Our spectra show bright lines 
[CaII] 7291, 7323 \AA, which suggests that a warm dust envelope surrounds the star. 
\cite{Kraus2014} also detected CO lines in the spectra. The CO lines are 
very weak (although detectable) in our spectra, which can be explained by variability 
of these lines. That was noticed before in the case of HR Car by \cite{Morris1997}.
From the presence of [CaII] and CO lines, and from the location at the 
B[e]SG region on the JHK diagram, \cite{Kraus2014} have concluded 
that this object is a B[e]SG star. \cite{Humphreys2014} classify it as a 
FeII-emission line star. Our results confirm these two conclusions.

Interesting that while B[e]SGs do not show significant spectral
variability, this star was variable during last 17 years. 
To date, only one B[e]SG, S18 \cite[]{Clark2013}, showed such a variability. 
Note however that classification of S18 as a B[e]SG is not ultimate yet. 
The spectral variability of J004417.10 requires further investigation. 

\noindent
{\it J004444.52+412804.0} \,
Variability of this star ($\Delta$B = 0.27, $\Delta$V = 0.22) was 
noticed in the DIRECT project \cite[]{Stanek1999}. \cite{Vilardell2006} 
confirmed the variability with amplitudes of $\Delta$B = 0.30 and 
$\Delta$V = 0.25 between 1999 and 2003. Their mean magnitudes correspond to 
those measured by us (Table 2). Our spectrum is almost identical to 
that by \cite{Massey2007} and similar to that by \cite{Humphreys2013,Humphreys2014}. 
The spectra have emission lines FeII and [FeII]. Helium lines 5876~\AA\, 
and 6678~\AA\, have P Cyg profiles. We also observe bright emission lines 
[CaII] 7291, 7323~\AA\,, and $^{12}$CO. \cite{Humphreys2013} conclude 
that the star is a warm supergiant, but their parameters $A_V=1.5 \div 2.6$ mag, 
T = 7000 $\div$ 9000 K (correspond to F0Ia spectral class) do not agree
with our estimates from SED. 
This spectral class also does not agree with the presence of 
HeI emission in our spectra, same to the spectra by \cite{Massey2007}.
The spectrum taken by \cite{Humphreys2013} also shows He\,I line, 
however, it also has signatures of O\,I~$\lambda 7774$, Ca\,II H and K, 
Ti\,II blends in absorption. The latter means a lower gas temperature.
Probably, the star has very extended atmosphere with a lower wind
ionization in outer parts. 

The star locates at the B[e]SG region on the JHK diagram. 
This is the second bright star (together with J004417.10) located in 
the B[e]SG region \citep{Kraus2014} that shows a significant brightness 
variability. 
We classify this star as a B[e]SG. However, its optical variability  
is marginal for LBVs ($\approx 0.3$ mag). A long-term photometric
monitoring of the star is needed to confirm its nature. 

\noindent
{\it J004526.62+415006.3}\, 
\cite{Vilardell2006} identified this object as a variable star. Our 
data suggest its clear LBV-like variability: the star reddens when the brightens
(Fig. 4 and Table 2). The photometric variability ($\Delta V = 1.0$ mag)  is 
followed by the spectral one: the spectra by \cite{Massey2007} show numerous FeII
emission, weak HeI emission, whereas in our spectra FeII got 
much weaker or even turned to the absorption and no HeI was detected.  
Our spectrum is cooler than Massey's: our SED suggests T=12000~K, 
while fitting SED to that from \cite{Massey2007} gives T=18300~K, 
given  $A_V$ = 1.3 $\pm$ 0.1. The photometric and spectroscopic 
variability allow us to classify this object as an LBV. 

\cite{Humphreys2014} noticed that the star's spectrum closely 
resembles that of J004444.52. However, they did not have information
about $\sim 1$ mag LBV-like variability, which we demonstrate in this paper. 
This gives us a sign that J004444.52 might also show a strong variability in 
the future.

\noindent
{\it Var\,A-1} 
Our spectra show broad Balmer lines, bright HeI emission, and 
numerous strong FeII and [FeII] lines. 
The spectrum obtained by us is similar to that 
published by \cite{Humphreys2014}. 
We use three data sets to fit SEDs: 
that by \cite{Humphreys1984} obtained in 1976, ours described 
in this paper, and photometry by \cite{Massey2006} from 2000 - 2001.
The SED fitting suggests T=20,400 K for the maximum brightness 
by \cite{Humphreys1984}, for the intermediate brightness
(our data) we find T=25,000K, and T=28,100K for the minimum 
\cite[]{Massey2006}. The estimated temperatures agree with 
the observed spectral features. The reddening $A_V$ = 1.8 mag 
estimated for the nearby nebula corresponds to our SED estimate 
($A_V = 1.7\pm0.1$ mag). 

\noindent
{\it AE And} 
Our spectra show broad Balmer lines, and numerous and strong FeII and 
[FeII] lines. 
Our spectrum is similar to that shown by \cite{Humphreys2014}.  
We also identify [NiII] 6668, 6813~\AA\, lines, in our spectra,
the same lines we find in the spectra published by \cite{Szeifert1996}.
In the contrast to the spectra by \cite{Massey2007}, the [FeIII] and [NII] 5755~\AA\, 
lines disappear, FeII lines got weaker, and HeI lines are barely seen 
in our spectra. Due to a good number of photometric data points and the strong 
variability of the star, we are enabled to employ the data for three 
well separated in time epochs: our data, photometry by \cite{Massey2006} (2000 - 2001), 
and photometry by \cite{Humphreys1984} (1976). 
The obtained temperatures are in a good agreement with the brightness 
and the spectral features seen in each data set.  
Our spectra indicate the coolest temperature (T = 18,000 K), the hottest 
temperature is found for Massey's data (T = 25,400 K), and the intermediate 
temperature is found for Humphreys's data (T = 21,600 K). We find 
$A_V = 1.0\pm0.1$ mag. 
Since our analysis is based on the data sets corresponding to the 
different states of the LBV stars, we believe that the parameters derived by 
us are more reliable.

\begin{table*}
\begin{center}
\caption{Results of our SED modeling. The columns show the object name, the 
%guessed 
photosphere temperature range preliminary estimated from spectra,
the best-fit temperature, the reddening, the stellar radius in the solar units,  
the V-band and bolometric absolute magnitudes.
}
\begin{tabular}{ccccccc}
\hline
Object & $T_{sp}$, K & $T_{SED}$, K& $A_{V}$, mag & $R/R_{\odot}$ &$M_V$, mag &$M_{bol}$, mag \\% & $T_{dust}$, K
\hline
J004051.59 & 18000 - 24000 & 22000$\pm$2000 &1.5$\pm$0.1 & 90 & -9.0 & -10.9$\pm$0.2 \\%  & 1500  \\
J004350.50 & 10000 - 15000 & 13000$\pm$2500 &2.0$\pm$0.2 &130 & -8.7 & -9.4$\pm$0.2 \\%  & 1500   \\
J004417.10 & 15000 - 20000 & 18000$\pm$1000 & 1.0$\pm$0.2 &70& -8.1 & -9.6$\pm$0.1 \\ % & 1500  \\
J004444.52 & 15000 - 20000 & 18000$\pm$2000 & 3.6$\pm$0.1 & 160& -9.8 &-11.2$\pm$0.2 \\%  &1300  \\
J004526.62 & 10000 - 15000 & 12000$\pm$2000 &1.3$\pm$0.1 & 200& -9.4 & -10.0$\pm$0.2 \\%  & 1500  \\
Var\,A-1     & 20000 - 27000 & 25000$\pm$1000 & 1.7$\pm$0.1 & 90& -9.3 & -11.5$\pm$0.1 \\%  & 1500 \\
AE And     & 15000 - 20000 & 18000$\pm$1000 &1.0$\pm$0.2 & 100& -8.9 &-10.3$\pm$0.2  \\% & 1500 \\
\hline
\end{tabular}
\end{center}
\end{table*}

\section{Conclusions}
We develop a new method of SED fitting applicable for LBV stars. It breaks a well-known 
reddening-stellar temperature degeneracy 
by an additional assumption 
that the bolometric luminosity stays constant while the optical V-brightness
may vary significantly. This is an inherent property of LBVs 
to change their spectral type keeping its bolometric luminosity constant. 
Having the bolometric luminosity and the reddening fixed, we calculate 
the V-magnitudes and the stellar temperatures for different states of LBVs. 
Our approach is successfully verified with two known LBVs in the Andromeda 
galaxy: Var\,A-1 and AE And.

All considered 
LBV candidates show spectra typical for either LBVs or B[e]SGs. The 
luminosity ranges are in agreement with those for known  LBVs in M31. 
All the stars show brightness variability, both in ours and literature data.  
In two stars, J004417.10 and J004526.62, 
we have also detected spectral variability. We conclude that two studied 
stars, J004051.59 and J004526.62, are LBVs, 
because they show significant brightness variability. 
Since J004350.50 does not show sufficient variability, 
we can not classify the star confidenly, although in its spectrum, 
luminosity and location in the JHK diagram the star is similar to
LBVs. The star does not show an obvious LBV-like variability and 
therefore, it remains an LBV-candidate. One needs to use additional 
criteria (e.g. helium content) to clarify its nature.   

We identify two stars J004417.10 and J004444.5 as B[e]-supergiants. They 
have an excess in the JHK bands, especially prominent in the K. 
They are located in the B[e]SG region on the JHK diagram \cite[]{Kraus2014}. 
Nevertheless, both stars indicated variability, J004417.10 
changed its spectrum, whereas J004444.5 changed its optical brightness. 
Because of the variability we do not exclude that the B[e]SG classification 
of these stars can be changed in the future. 

\section{Acknowledgments}

The research was supported by the Russian Scientific Foundation (grant 
N\,14-50-00043), is partly supported by RFBR grants N\,13-02-00885, ``Leading
Scientific Schools of Russia'' 2043.2014.2. D.B. was partly supported by
RSF grant RSCF-14-22-00041. S.F. acknowledges support of
the Russian Government Program of Competitive Growth of Kazan Federal University.
Based on observations obtained with the Apache Point Observatory 3.5-meter telescope, 
which is owned and operated by the Astrophysical Research Consortium. 
We thank our referee, K. Davidson, whose 
constructive comments and suggestions significantly improved the paper.

\label{lastpage}
\end{document}